\documentclass[journal]{IEEEtran}
\usepackage{amsmath}
\usepackage{graphicx}
\usepackage{wasysym}
\usepackage[unicode=true]
 {hyperref}

\makeatletter
\let\oldforeign@language\foreign@language
\DeclareRobustCommand{\foreign@language}[1]{%
  \lowercase{\oldforeign@language{#1}}}

\ifCLASSINFOpdf
\else
\fi
\usepackage{xcolor}
\newcommand\ed[1]{{\color{black}{#1}}}
\newcommand\revtwo[1]{{\color{black}{#1}}}
\newcommand\Amat{{\textbf{\textit{A}}}}
\renewcommand\v[1]{{\textbf{\textit{#1}}}}

\usepackage{acronym}
\acrodef{DUT}[DUT]{device under test}
\acrodef{VNA}[VNA]{vector network analyzer}
\acrodef{NIST}[NIST]{National Institute of Standards and Technology}

\makeatother

\begin{document}
\title{Measuring Noise Parameters Using an Open, Short, Load, and $\lambda$/8-length
Cable as Source Impedances}
\author{D. C. Price, C. Y. E. Tong,~\IEEEmembership{Member,~IEEE,} A. T. Sutinjo,~\IEEEmembership{Senior Member,~IEEE,}
L. J. Greenhill, N.Patra
\thanks{D. C. Price, A. Sutinjo and N. Patra are with the International Centre
for Radio Astronomy Research, Curtin University, Bentley WA 6102,
Australia.}
\thanks{L. J. Greenhill and E. Tong are with  Center for
Astrophysics, Harvard \& Smithsonian, 60 Garden Street, Cambridge, MA 02143, USA}
\thanks{Manuscript received Oct xx 2021; revised mmm dd 20yy.}}
\markboth{IEEE TRANSACTIONS ON MICROWAVE THEORY AND TECHNIQUES}{Authors \MakeLowercase{et al.}: A singularity-free noise parameter
measurement method}
\maketitle
\begin{abstract}
Noise parameters are a set of four measurable quantities which determine
the noise performance of a radio-frequency device under test. The
noise parameters of a 2-port device can be extracted by connecting
a set of 4 or more source impedances at the device's input, measuring
the noise power of the device with each source connected, and then
solving a matrix equation. However, sources with high reflection coefficients
($\left|\Gamma\right|\approx1$) cannot be used due to a singularity
that arises in entries of the matrix. Here, we detail a new method
of noise parameter extraction using a singularity-free matrix that
is compatible with high-reflection sources. We show that open, \ed{short},
load and an open cable (``OSLC'') can be used to extract noise parameters,
and detail a practical measurement approach. The OSLC approach is
particularly well-suited \revtwo{for low-noise amplifiers} at frequencies below 1 GHz, where alternative
methods require physically large apparatus.
\end{abstract}



\IEEEpeerreviewmaketitle{}
\bstctlcite{IEEEexample:BSTcontrol}

\section{Introduction}

\IEEEPARstart{T}{he} noise performance of a radio-frequency amplifier,
or other \ac{DUT}, is commonly characterized in terms
of its noise parameters: a set of four \ed{real-valued} terms from which noise characteristics
can be derived for all input and output impedances \cite{Haus:1960}.
Alternatively\textemdash but equivalently\textemdash a ``noise wave''
representation may be used, which defines noise in terms of incoming
and outcoming waves \cite{Meys:1978,Wedge:1992}. Regardless of representation,
the measurement of noise parameters is an important task when determining
and optimizing the signal-to-noise performance of a radio receiver. 

\ed{This article, we present two main results. Firstly, we reintroduce and expand on a matrix formulation for determining noise parameters, which allows for sources with 
$\left|\Gamma_{s}\right|\approx1$ to be used. Central to this approach is a 
singularity-free matrix formed from the reflection coefficients of four sources. 
We show the relationship between the singularity-free matrix and the traditional admittance-based
matrix formulation \cite{Lane:1969}, then show that the singularity-free formulation yields smaller measurement errors. 
Compared to standard techniques, no change in measurement apparatus is required;
as such our approach can be used a substitute for
the admittance-based matrix formulation.}

\ed{Secondly, we detail a cold-source technique for measurement
of noise parameters based upon the singularity-free matrix formulation. Our measurement technique requires only a 1/8-wavelength
coaxial cable and open, short and load termination.  
A key feature of this technique is the use of open and short source impedances, for which well-characterized commercial offerings are readily available as part of precision \ac{VNA} calibration kits.
The technique can be used 
with any unconditionally stable \ac{DUT}, and is ideally suited to low-frequency application ($<$1\,GHz). 
}

This paper is organized as follows. \ed{We first give an overview of noise parameters 
 (Sec. \ref{sec:noise-params}) and matrix-based
approaches (Sec. \ref{sec:Matrix-based-noise-parameter}), then introduce
a singularity-free matrix formulation (Sec. \ref{sec:A-reflection-coefficient})}.
We then outline how noise parameters can be measured using an open,
short, load and 1/8-wavelength coaxial cable as reference source impedances
 (Sec. \ref{sec:Using-open-and} and \ref{sec:method}). In Sec. \ref{sec:Measurement-example},
we use our approach to measure the noise parameters of a Minicircuits
ZX60-3018G-S+ amplifier across 50\textendash 300\,MHz. The paper
finishes with a discussion and concluding remarks (Sec. \ref{sec:Discussion}).

\ed{\section{Noise parameters}\label{sec:noise-params}}

\begin{figure}
\begin{centering}
\includegraphics[width=1\columnwidth]{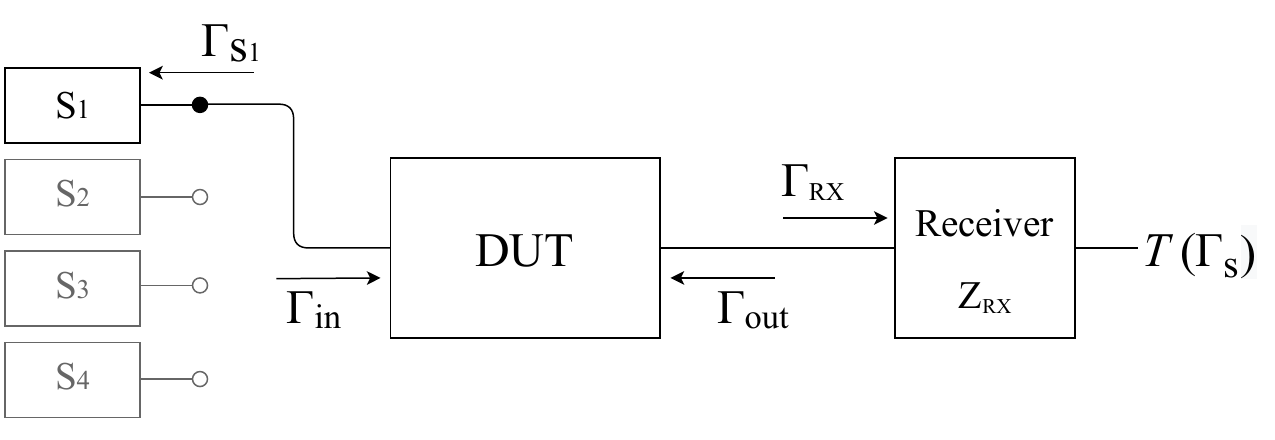}
\par\end{centering}
\caption{Diagram showing setup of a device-under-test (DUT) for noise parameter
measurements using the cold-source method. A set of four (or more)
reference source impedances $\Gamma_{s}$ are connected to the DUT
and $T(\Gamma_{s})$ is measured with a noise receiver (RX). The noise
figure of the DUT must first be measured for a single source impedance,
by using a calibrated noise source. \label{fig:dut-diagram}}
\end{figure}

In terms of source reflection coefficient $\Gamma_{s}$, or source
admittance $Y_{s}=G_{s}+jB_{s},$ the noise temperature $T$ of a
2-port DUT can be expressed as :
\begin{align}
T(\Gamma_{s}) & =T_{{\rm min}}+T_{0}\frac{4R_{N}}{Z_{0}}\frac{\left|\Gamma_{s}-\Gamma_{{\rm opt}}\right|^{2}}{(1-\left|\Gamma_{s}\right|^{2})\left|1+\Gamma_{{\rm opt}}\right|^{2}}\label{eq:noise-param}\\
T(Y_{s}) & =T_{{\rm min}}+T_{0}\frac{R_{N}}{G_{s}}\left|Y_{s}-Y_{{\rm opt}}\right|^{2}\\
T(G_{s},B_{s}) & =T_{{\rm min}}+T_{0}\frac{R_{N}}{G_{s}}\left[(G_{s}-G_{{\rm opt}})^{2}+(B_{s}-B_{{\rm opt}})^{2}\right]
\end{align}
where $T_{0}=290$ K and $Z_{0}=1/Y_{0}$ is the characteristic impedance.
$T$ is comprised of \ed{the following} noise parameters:
\begin{itemize}
\item $T_{{\rm min}}$ is the minimum noise temperature, also commonly expressed
as the minimum noise factor, $F_{{\rm min}}=(1+T_{{\rm min}}/T_{0})$
\item $Y_{{\rm opt}}=G_{{\rm opt}}+jB_{{\rm opt}}$ is the optimum admittance,
or equivalently, $\Gamma_{{\rm opt}}=\gamma_{{\rm opt}}\,exp\left(j\theta_{{\rm opt}}\right)$
is the optimum reflection coefficient. 
\item $R_{N}$ is the equivalent noise resistance. Alternatively, the unitless
quantity $N=R_{N}G_{{\rm opt}}$ may be used, which is invariant under
reciprocal lossless transformations.
\end{itemize}
There are several approaches to extract noise parameters from measurements
of the noise temperature $T(\Gamma_{s})$. In all approaches, as there
are four \ed{unknown (real-valued)} noise parameters, at least \ed{$n$$\ge$4} independent measurements
of $T(\Gamma_{s})$ must be made. The noise parameters are then found
by casting the problem as a matrix equation (see Sec.\,\ref{sec:Matrix-based-noise-parameter}) or by
equivalent least-squares methods. The loci of the $n$ reference $\Gamma_{s_{i}}$
on the Smith chart will form an ``impedance pattern'', and it has
long been recognized that loci should be ``well spread'' across
the Smith chart \cite{Davidson:1989,VanDenBosch:1998,DeDominicis:2002,Himmelfarb:2016}.
In general, a vector network analyzer (VNA) is used to accurately
measure $\Gamma_{s}$, and a noise receiver is used to measure $T(\Gamma_{s})$,
see Fig. \ref{fig:dut-diagram}. 

While any four (or more) source impedances may be used, impedance
tuners are commonly employed as they offer a convenient way to generate
well-spaced impedances. We will refer to the case in which four source
impedances are used, with a load ($\Gamma=0$) selected as one source
impedance, as the ``four-point'' method. An alternative approach
is to use a source impedance which exhibits rapid phase wraps across
frequency, such as a long coaxial cable \cite{Hu:2004,rogers:2012}.
Under the assumption that the noise perfomance of the DUT does not
change appreciably across a small frequency range, the phase wrapping
may be used to effectively sample a range of impedances across the
Smith chart. 

Regardless of approach, a well-characterized noise source is required
to calibrate receiver noise power measurements (using the Y-factor
method \cite{Friis:1944}). By inserting a two-port impedance tuner
between the noise source and the DUT, the noise figure can be measured
for different source impedances, as required for noise parameter extraction.
An alternative ``cold-source'' technique can also be used, in which
the noise source is directly connected to the DUT and the noise figure
is measured, after which a set of passive source impedances is connected
to provide measurements at different source impedances \cite{Adamian:1973}.
In the cold-source method, the noise figure of the DUT is measured
for only one source impedance\textemdash that of the noise source\textemdash and
one-port devices are used as source impedances.

All variations of these approaches require that reflection coefficients
are not too large, due to a singularity (division by zero) caused
by the $(1-\left|\Gamma_{s}\right|^{2})$ term in Equation \ref{eq:noise-param}.
Himmelfarb \& Belostotski (henceforth HB16) \cite{Himmelfarb:2016}
provides a mathematical basis to show that one source impedance should
be a 50$\Omega$ load, and three (or more) reflection coefficients
should satisfy $0.4<\left|\Gamma_{s_{i}}\right|<0.9$. By doing so,
the well-spaced requirement is met, and the $(1-\left|\Gamma_{s}\right|^{2})$
singularity is not encountered. 

At low frequencies ($<$100 MHz), the long cable approach is troublesome,
as the cable can become prohibitively long. For example, an application
to measure noise parameters at 50\textendash 200\,MHz for radio astronomy
suggests the use of a 25\,m cable \cite{Roque:2021}. Low-frequency
impedance tuners are also physically large, and can be prohibitively
expensive. 

\section{Matrix-based noise parameter approaches\label{sec:Matrix-based-noise-parameter}}

The extraction of noise parameters from a DUT requires connecting
$\ed{n}$$\ge$4 reference sources and measuring the noise output power
spectra of the DUT using a receiver. The reflection coefficients $\Gamma_{s}$,
or equivalent admittances $Y_{s}$, must be known or measured for
each source. In matrix notation, the noise receiver measurements form
a ($n\times$1) vector \v{t}, the $\Gamma_{s}$ measurements
form a ($n$$\times$\ed{4}) matrix \ed{\Amat}, and we wish to find the (4$\times$1)
noise parameter vector \v{t}, which is related by: 
\begin{equation}
\Amat\v{x}=\v{t}.
\end{equation}
To solve this (i.e. find \v{t}) requires inverting the matrix
$\Amat^{-1}$ (if $n$=4) or forming the pseudoinverse $\Amat^{+}=(\Amat^{T}\Amat)^{-1}\Amat^{T}$
if $n>$4:
\begin{equation}
\v{x}^{+}=\Amat^{+}\v{t}.\label{eq:pseudoinverse}
\end{equation}
The entries of the matrix \Amat\, depend upon the formulation used, of
which there are a several. In Lane's technique \cite{Lane:1969},
the $i$th row of \Amat\, is formed from admittances:

\begin{align}
{\Amat}_{i}^{G} & =\left[1,\frac{\left|Y_{s_{i}}\right|^{2}}{G_{s_{i}}},\frac{1}{G_{s_{i}}},\frac{B_{s_{i}}}{G_{s_{i}}}\right],
\end{align}
and the vector $\v{x}=[a,b,c,d]^{T}$, whose entries can be converted
into the four noise parameters by:
\begin{align}
T_{{\rm min}} & =a+\sqrt{4bc-d^{2}}\\
R_{N} & =b\\
G_{{\rm opt}} & =\sqrt{4bc-d^{2}}/2b\\
B_{{\rm opt}} & =-d/2b.
\end{align}
An alternative formulation is found in \cite{Hu:2004}, which defines
a matrix in terms of the magnitude $\gamma_{s}$ and phase $\theta_{s}$
of $\Gamma_{s}=\gamma_{s}exp(j\theta_{s})$
\begin{align}
{\Amat}_{i}^{\gamma} & =\left[1,\frac{1}{1-\gamma_{i}^{2}},\frac{\gamma_{i}cos\theta_{i}}{1-\gamma_{i}^{2}},\frac{\gamma_{i}sin\theta_{i}}{1-\gamma_{i}^{2}}\right],
\end{align}
and noise parameters\footnote{Note the minus signs in Equation \ref{eq:hu-np-theta} \textendash{}
omitted in \cite{Hu:2004} \textendash{} are important to ensure the
correct quadrant is returned when using $tan^{-1}$. } are obtained as
\begin{align}
T_{{\rm min}} & =a+\frac{b+\Delta}{2}\label{eq:hu-np-Fmin}\\
R_{N} & =\frac{\Delta}{4Y_{0}}\\
\gamma_{{\rm opt}} & =\sqrt{\frac{b-\Delta}{b+\Delta}}\\
\theta_{{\rm opt}} & =tan^{-1}\left(\frac{-d}{-c}\right),\label{eq:hu-np-theta}
\end{align}
where $\Delta=\sqrt{b^{2}-c^{2}-d^{2}}$. 

\section{A reflection coefficient based source matrix\label{sec:A-reflection-coefficient}}

\ed{When measuring noise parameters, the choice of source impedances is crucial to minimize
measurement error.
In this section, we introduce the matrix $\Amat^\Gamma$, and show that
by using $\Amat^\Gamma$ in lieu of $\Amat^\gamma$ or $\Amat^G$ yields lower measurement errors.}

The invertibility of matrices $\Amat^{\gamma}$ and $\Amat^{G}$ depends upon
the characteristics of the reference sources. Of the $n$ reference
loci, one is almost always chosen to be the $\Gamma=0$ reference
impedance. A singularity is encountered in the entries $\Amat^{\gamma}$
and ${\Amat}^{G}$ if $\left|\Gamma_{s}\right|^{2}\rightarrow1$, so open
and short references cannot be used. 

In Sutinjo et. al (henceforth SUT20) \cite{Sutinjo:2020}, it is shown
that the singularities in $\Amat^{\gamma}$ and $\Amat^{G}$ can be removed
after multiplication by $(1-\left|\Gamma_{s}\right|^{2})$ \cite{Sutinjo:2019,Sutinjo:2020}.
From Equation \ref{eq:noise-param}, as $\left|\Gamma_{s}\right|\rightarrow1$
we see that $T(\Gamma_{s})\rightarrow\infty$ due to the $(1-\left|\Gamma_{s}\right|^{2})$
term in the denominator. However, in the limit $\left|\Gamma_{s}\right|\rightarrow1$,
we have 
\[
\lim_{|\Gamma_{s}|\to1}\left((1-\left|\Gamma_{s}\right|^{2})T(\Gamma_{s})\right)=T_{0}\frac{4R_{N}}{Z_{0}}\frac{\left|\Gamma_{s}-\Gamma_{{\rm opt}}\right|^{2}}{\left|1+\Gamma_{{\rm opt}}\right|^{2}};
\]
that is, the quantity $(1-\left|\Gamma_{s}\right|^{2})T(\Gamma_{s})$
is non-zero. The two matrices thus become:
\begin{align}
{\Amat}_{i}^{G'} & =(1-\left|\Gamma_{s_{i}}\right|^{2})\left[1,\frac{\left|Y_{s_{i}}\right|^{2}}{G_{s_{i}}},\frac{1}{G_{s_{i}}},\frac{B_{s_{i}}}{G_{s_{i}}}\right]\\
{\Amat}_{i}^{\gamma'} & =\left[1-\gamma_{i}^{2},1,\gamma_{i}cos\theta_{i},\gamma_{i}sin\theta_{i}\right].
\end{align}
By removing the singularity, SUT20 provided a physical and mathematical
basis for why loci in the impedance pattern should be ``well spread''.
For $n=4$ measurements, the matrix $\Amat'$ is $4\times4$ and the maximum
spread on the Smith chart corresponds to maximizing the magnitude
of the matrix determinant $|$det$|$. SUT20 also show that the condition
number of the matrix $\Amat'$ is strongly anti-correlated, and is minimized
for maximum $|$det$|$; in contrast, for un-regularized matrices, i.e.
$\Amat^{G}$ and $\Amat^{\gamma}$, the condition number  is an unreliable
predictor \cite{VanDenBosch:1998}. 

Here, we highlight that $\Amat^{G'}$can be rewritten as
\begin{align}
{\Amat}_{i}^{G'} & =\left[1-\left|\Gamma_{s_{i}}\right|^{2},\,\left|1-\Gamma_{s_{i}}\right|^{2},\,\left|1+\Gamma_{s_{i}}\right|^{2},-2{\rm Im}(\Gamma_{s_{i}})\right]\equiv {\Amat}_{i}^{\Gamma}.\label{eq:AGamma}
\end{align}
A derivation of this result is provided in Appendix \ref{sec:Derivation-of}.
The matrix $\Amat^{\Gamma}$, is numerically equivalent to $\Amat^{G'}$,
but its entries are simple expressions of $\Gamma_{s}$. As such,
there is no need to convert source reflection coefficients into admittances.
Using $\Amat^\Gamma$, the matrix relation becomes
\begin{align}
\Amat^{\Gamma}\v{x} & =\v{t}'\label{eq:main-matrix-relation}\\
\v{x} & =[a,b,c,d]^{T}\label{eq:main-matrix-vector}\\
\v{t}' & =(1-\left|\Gamma_{s_{i}}\right|^{2})\v{t}\label{eq:main-matrix-mod}
\end{align}
where the noise parameters are related to the $\v{x}=[a,b,c,d]^{T}$
vector by:
\begin{align}
T_{{\rm min}} & =a+\sqrt{4bc-d^{2}}\label{eq:np-fmin}\\
R_{N} & =b/(Y_{0}T_{0})\\
G_{{\rm opt}} & =Y_{0}\sqrt{4bc-d^{2}}/2b\\
B_{{\rm opt}} & =-Y_{0}d/2b.\label{eq:np-bopt}
\end{align}
The reverse relations are provided in Appendix \ref{sec:Reverse-relations}.

\ed{The source impedance matrix $\Amat^\Gamma$ (Eq.\,\ref{eq:AGamma}) is a central 
result of this paper. 
It can be employed in any noise parameter extraction technique
based upon $\Amat^G$ by minor modification
to the matrix relation (Eqs.\ref{eq:main-matrix-relation}--\ref{eq:main-matrix-mod}). 
Similarly, the matrix $\Amat^\gamma{'}$ can be used in lieu of $\Amat^\gamma$.
We will now show that these substitutions are well motivated as they minimize 
errors arising from matrix inversion.}

\section{Comparison of measurement errors \label{subsec:Matrix-inversion-errors}}

\ed{The propagation of errors in matrix inversion problems is non-trivial, particularly if
row entries are covariant \cite{Lefebvre:2000}. Within each row of \Amat, entries are indeed 
highly covariant as they are formed from the same source admittance/reflection measurement. Nevertheless, simple arguments
about worst-case scaling errors (i.e. propagation of relative
errors due to matrix inversion) can be made based upon the matrix determinant.
} 

It has been previously noted that higher $|$det$|$ for \Amat\ corresponds to lower absolute
uncertainties \cite{DeDominicis:2002}. SUT20 show that the maximum
possible $|$det$|$ for $\Amat^{G'}$ (and thus $\Amat^{\Gamma}$) is 41.57, when
points are maximally spread; for the $|\Gamma_{s}|<0.9$ 
case (required for $\Amat^G$) the maximum $|$det$|$ $<27.7$. 
\ed{This suggests that $\Amat^{\Gamma}$ has the potential to yield lower 
worst-case scaling errors by allowing the use of reference sources with $|\Gamma_{s}|>0.9$.}

\ed{Seemingly counter to this,} HB16 argues that high-reflection sources 
intrinsically introduce measurement
uncertainties separate to scaling errors due to matrix inversion.
To illustrate, they use concentric noise circles of 0.1-dB around
$\Gamma_{{\rm opt}},$which when plotted on the Smith chart become
denser toward the edge. However, by scaling measurements by $(1-|\Gamma_{s}|)^{2}$,
the space between concentric rings is constant, negating this effect. 

In four-point methods, several authors have noted that measurement
error is minimized when three of the source impedances are purely
real (i.e. the phase of $\Gamma$ is $0^{\circ}$, or $180^{\circ}$),
and one impedance is located at $90^{\circ}$ or $270^{\circ}$ on
the Smith chart \cite{VanDenBosch:1998,DeDominicis:2002,Belostotski:2010}.
HB16 explains that this corresponds to forming a matrix of source
impedances that is diagonally dominant.

A previous study determined that errors do not increase meaningfully as long as a minimum spacing
of 30 degrees between points in the impedance pattern is maintained
\cite{DeDominicis:2002}. \ed{This requirement is equivalent to
setting a minimum acceptable $|$det$|$; for $\Amat^\Gamma$, the requirement
corresponds to $|$det$|\apprge$ 15.5. Using a similar argument, SUT20 suggests $|$det$|$ $\apprge$
10.}

\ed{To summarize, the selection of reference sources affects the errors we expect due to
matrix inversion. Errors are minimized by maximizing $|$det$|$, and by choosing sources that 
correspond to a diagonally dominant matrix. We now demonstrate that for a given four-point pattern,
measurement errors can be improved just by choosing to use $\Amat^\Gamma$ in lieu of $\Amat^G$. 
}

To qualitatively compare how the choice of matrix {\Amat} affects errors,
we ran a Monte Carlo simulation of a ``toy'' DUT with $T_{{\rm min}}=200$
K, $\left|\Gamma_{{\rm opt}}\right|=0.3$, $\theta_{{\rm opt}}=90^{\circ}$
and $N$=0.25. \ed{These values were chosen to be similar to those measured in Sec.\,\ref{sec:Measurement-example} for a Minicircuits ZX60-3018G-S+ amplifier.} 
Using the relations Appendix \ref{sec:Reverse-relations},
we converted noise parameters into a vector $\v{x}$, and then
computed the measurement vector $\v{t}=\Amat\v{x}$ for both matrices
$\Amat^{G}$ and $\Amat^{\Gamma}$. The \Amat\ matrices were formed using an
four-point pattern $\Gamma_{s_{i}}=(0,0.9,-0.9,0.9\angle90^{\circ})$, which has maximum spacing
between loci on the Smith chart.

\begin{figure}
\centering{}\includegraphics[width=0.99\columnwidth]{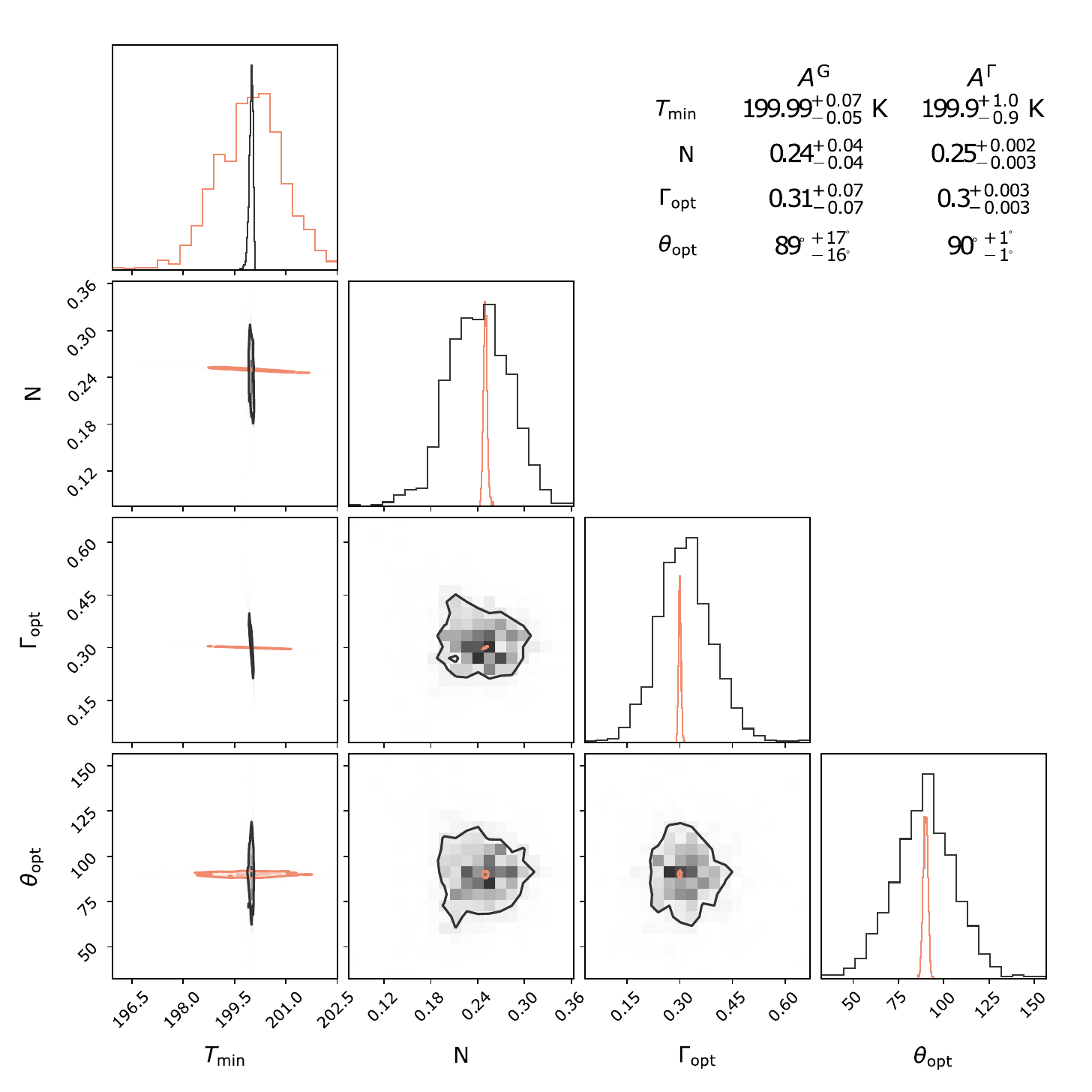}\caption{\ed{Scatterplot matrix showing noise parameter uncertainties for a toy model DUT. Each scatter plot within the matrix visualizes the covariance relationship between a pair of parameters. Points computed using the traditional admittance matrix $\Amat^{G}$ are shown in black, and the reflection-coefficient matrix $\Amat^{\Gamma}$
in red.}  \label{fig:old-mat-vs-new-mat}}
\end{figure}

To simulate errors in $\Gamma_{s}$, we generated normally-distributed
0.1\,dB magnitude and $1^{\circ}$ phase offsets, then computed error-corrupted
noise vectors $\tilde{\v{x}}$ via:
\begin{equation}
\tilde{\v{x}}=\left(\tilde{\Amat}\right)^{-1}\v{t},
\end{equation}
where $\tilde{\Amat}$ is the source matrix with errors. 
\ed{This approach is similar to that used by \ac{NIST} for noise parameter uncertainty analysis \cite{Randa:2015}.} We ran this procedure
1024 times, then formed scatterplot matrices of noise parameter estimates
(Figure \ref{fig:old-mat-vs-new-mat}); \ed{the scatterplot shows the covariance between parameters}. 
We find that the traditional approach with $\Amat^{G}$ exhibits order-of-magnitude larger uncertainties
in the retrieved values for $N$, $|\Gamma_{{\rm opt}}|$ and $\theta_{{\rm opt}}$,
but smaller uncertainties on $T_{{\rm min}}$ (i.e. if covariance
is ignored). We conclude that the removal of the singularity as $\left|\Gamma_{s}\right|\rightarrow1$
does indeed lead to smaller errors due to matrix inversion.

\section{Using open and short source references\label{sec:Using-open-and}}

\ed{In the previous section, we showed that for a given four-point pattern, using $\Amat^\Gamma$ in lieu of $\Amat^G$ leads to smaller errors. Here, we discuss the use of four-point patterns comprised of highly-reflective sources, which cannot be used with $\Amat^G$.}

\ed{The use of open and short impedances ($\Gamma_{{\rm op}}=1$ and $\Gamma_{{\rm sh}}=-1$ ) is well motivated by previous four-point studies, which found that measurement errors are minimized when the phase of $\Gamma_{s}$ is $0^{\circ}$, or
$180^{\circ}$ for three of the impedances \cite{VanDenBosch:1998,DeDominicis:2002,Himmelfarb:2016}. In particular, HB16 identifies four regions on the Smith chart in which measurements should be made: regions ``B'' and ``C'' contain open and short loci, but the $|\Gamma|<0.9$ requirement precluded their use.

While using open and short maximizes $|$det$|$ of \Amat, leading to smaller matrix inversion errors, we must also consider the measurement accuracy for $\Gamma_s$. In general, fractional $S_{11}$ measurement uncertainties on a given \ac{VNA} are greater for highly-reflective sources; this is a separate error term that runs counter to maximizing $|$det$|$. Fortunately, physical models of short and open are provided by calibration kit manufacturers, and methods for precise and accurate measurement of these standards are well understood \cite{dunsmore2011vna,Monsalve:2016}. As such, physical models of open and short sources can be used to characterize $\Gamma_{\rm{open}}$ and $\Gamma_{\rm{short}}$, to precision exceeding that possible by \ac{VNA} measurement alone.}
 
\ed{Let us consider} the case where ideal open, short and load ($\Gamma_{{\rm ld}}=0$)
are used as reference sources, along with an open 1/8-wavelength cable.
A lossless open (or shorted) 1/8-wavelength cable introduces a $90^{\circ}$
phase shift (or $-90^{\circ}$), such that $\Gamma_{{\rm cbl}}=\pm1j$. 

For these four references, $\Amat^{\Gamma}$ is
\begin{equation}
\Amat^{\Gamma}=\left(\begin{array}{c}
A_{1}^{\Gamma_{{\rm ld}}}\\
A_{2}^{\Gamma_{{\rm op}}}\\
A_{3}^{\Gamma_{{\rm sht}}}\\
A_{4}^{\Gamma_{{\rm cbl}}}
\end{array}\right)=\left(\begin{array}{cccc}
1 & 1 & 1 & 0\\
0 & 0 & 4 & 0\\
0 & 4 & 0 & 0\\
0 & 2 & 2 & \mp2
\end{array}\right).
\end{equation}
Note the entry $A_{44}=-2$ if an open cable is used, or $+2$ if
a shorted cable is used. $\Amat^{\Gamma}$ is invertible, with $|$det$|$=32
and condition number $c_{A}=5.62$. The inverse $\text{(\ensuremath{\Amat^{\Gamma}}})^{-1}$
\begin{equation}
\text{(\ensuremath{\Amat^{\Gamma}}})^{-1}=\frac{1}{4}\left(\begin{array}{cccc}
4 & -1 & -1 & 0\\
0 & 0 & 1 & 0\\
0 & 1 & 0 & 0\\
0 & 1 & 1 & \mp2
\end{array}\right).\label{eq:inverse}
\end{equation}
From this, the solutions to $\v{x}=\Amat\v{t}'$ are 
\begin{align}
a & =t'_{{\rm ld}}-(t'_{{\rm op}}+'t_{{\rm sh}})/4,\\
b & =t'_{{\rm sh}}/4,\\
c & =t'_{{\rm op}}/4,\\
d & =(t'_{{\rm op}}+t'_{{\rm sh}})/4\mp t'_{{\rm cbl}}/2,
\end{align}
from which we note that 1) $b$ and $c$ are directly given by the
measurement of short and open standards, respectively; and 2) the
load measurement is only required to compute $T_{{\rm min}}$, as
the \Amat\ term only appears in Equation \ref{eq:np-fmin}.

The matrix in Equation \ref{eq:inverse} is only correct at a central
frequency $f_{0}=v_{c}/\lambda_{0}$. For a 1/8-wavelength cable at
a central frequency $f_{0}=v_{c}/\lambda_{0}$, we can enforce a minimum
$|$det$|$ for the $\Amat^{\Gamma}$ matrix, and find a fractional bandwidth
over which our $|$det$|$ requirement is satisfied. The phase of $\Gamma_{{\rm cbl}}$
is given by $\theta(f)=4\pi L/v_{c}f$, where $v_{c}$ is the velocity
factor of the cable. Setting $|$det$|$$_{{\rm min}}$=10, based on the
recommendation of SUT20, we compute a corresponding frequency range
$f_{{\rm low}}$ to $f_{{\rm high}}$ :
\begin{align}
f_{{\rm low}} & =0.2f_{0}\label{eq:freq_min}\\
f_{{\rm high}} & =1.8f_{0}.\label{eq:freq_max}
\end{align}
Or put another way, $f_{{\rm high}}=9f_{{\rm low}}$, covering a 9:1
band. For example, if $f_{0}=1$ GHz, then the range over which $|$det$|$
$>$10 is 0.2\textendash 1.8 GHz.

\section{An Open-Short-Load-Cable Measurement method\label{sec:method}}

In this section, we present a practical method for noise parameter
extraction via the use of a load, open, short, and an open 1/8-wavelength
cable. In this method, which we call ``OSLC'', we form a measurement
vector $\text{\v{t}'},$which includes a $(1-\left|\Gamma_{s}\right|^{2})$
term that naturally arises when measuring the output power of a 2-port
DUT. This term cancels out the singularities inherent in previous
methods that use matrices $\Amat^{\Gamma}$ and $\Amat^{G}$.

The OSLC method relies on the matrix $\Amat^{\Gamma}$ (introduced in
Section 3), and requires the following:
\begin{itemize}
\item \ed{A calibrated} noise source to generate ``hot'' and ``cold'' temperature references, $T_{{\rm hot}}$
and $T_{{\rm cold}}$. The reflection coefficients,
$\Gamma_{{\rm hot}}$ and $\Gamma_{{\rm cold}}$ must be known or
measured, and should satisfy $\Gamma_{{\rm hot}}\approx\Gamma_{{\rm cold}}$.
\item A broadband load, open, and short, with known or measured reflection
coefficients $\Gamma_{{\rm ld}}$, $\Gamma_{{\rm op}}$, and $\Gamma_{{\rm sht}}$.
Additionally, an open cable (or other mismatch device) with $\Gamma_{{\rm cbl}}\approx\pm1j$.
To minimize pickup of radio interference, we recommend that the cable
is placed inside an RF-shielded box with an SMA feedthru connection.
\item \ed{A radio receiver} to measure power spectral density (PSD) with linear response. The input reflection coefficient $\Gamma_{{\rm rx}}$ 
should be known or measured, and the receiver must have high reverse
isolation ($S_{12}S_{21}\approx0$) for the analog component before
the digitizer.
\item A 2-port VNA to measure the S-parameters of the DUT, and any unknown
reflection coefficients. However, If the DUT is highly directional
and well-matched to the receiver \textendash{} such that $\left|S_{12}S_{21}\Gamma_{rx}\right|\approx0$
\textendash{} only $S_{11}$ is required. 
\end{itemize}
To show how the OSLC approach may be used, let us start by considering
a power measurement made by a radio receiver. The power $P_{s}$ measured
at the output of a 2-port network with scattering matrix {[}$S${]},
connected to a source impedance $Z_{s}$ and load impedance $Z_{{\rm rx}}$
(see Figure 1), is given by 
\begin{equation}
P_{s}=\mathcal{D}_{{\rm rx}}k_{B}\Delta f\mathcal{G}_{{\rm rx}}\mathcal{G}_{{\rm DUT}}\left(T_{s}+T_{{\rm n}}+\frac{T_{{\rm rx}}}{\mathcal{G}_{{\rm DUT}}(Z_{s})}\right)\label{eq:power-basic}
\end{equation}
where $k_{B}$ is the Boltzmann constant, $\Delta f$ is the noise
equivalent bandwidth, $T_{n}$ is the noise temperature of the DUT
(when connected to $Z_{s}$ and $Z_{{\rm rx}}$), $T_{s}$ is the
noise temperature of the source ($T_{s}=T_{{\rm amb}}$ for passive
networks at ambient temperature), and $\mathcal{G}_{{\rm DUT}}=\mathcal{G}_{{\rm DUT}}(Z_{s})$
is the available gain of the DUT. Here, $\mathcal{D}_{{\rm rx}}$
encompasses all (unknown) digital conversion and gain factors within
the receiver, assumed to be linear. 

The available gain \cite{pozar2011microwave}, denoted here with $\mathcal{G}$, of the \ac{DUT} is given by 
\begin{equation}
\mathcal{G}_{{\rm DUT}}(Z_{s})=\frac{\left|S_{21}\right|^{2}\left(1-\left|\Gamma_{s}\right|^{2}\right)}{\left|1-S_{11}\Gamma_{s}\right|^{2}\left(1-\left|\Gamma_{{\rm out}}(Z_{s})\right|^{2}\right)}.
\end{equation}
When connected to the source impedance $Z_{s}$, the two-port network
will be mismatched with a reflection coefficient, $\Gamma_{{\rm out}}(Z_{s})$:
\begin{equation}
\Gamma_{{\rm out}}(Z_{s})=S_{22}+\frac{S_{12}S_{21}\Gamma_{s}}{1-S_{11}\Gamma_{s}}.
\end{equation}
For a calibrated noise source with low reflection coefficient, and
requiring $\Gamma_{{\rm hot}}\approx\Gamma_{{\rm cold}}$, we define
$\Gamma_{{\rm ns}}=(\Gamma_{{\rm hot}}+\Gamma_{{\rm cold}})/2$. It
follows that the ratio
\begin{equation}
\frac{P_{{\rm hot}}-P_{{\rm cold}}}{T_{{\rm hot}}-T_{{\rm cold}}}=\mathcal{G}_{{\rm rx}}\mathcal{G}_{{\rm DUT}}k_{B}\Delta\nu.
\end{equation}
\ed{Where $\mathcal{G}_{\rm{rx}}$ is the available gain of the receiver's analog components.} Referring to Figure \ref{fig:dut-diagram}, the receiver sees an input
impedance $\Gamma_{{\rm DUT}}=\Gamma_{{\rm out}}(Z_{s})$, which depends
upon the source impedance $Z_{s}$. So, the total cascaded gain (as seen at the receiver output) is
given by $\mathcal{G}_{{\rm casc}}(Z)=\mathcal{G}_{{\rm DUT}}(Z)\mathcal{G}_{{\rm rx}}(Z_{{\rm out}})$.
For a receiver (with high reverse isolation), the ratio 
\begin{equation}
\frac{\mathcal{G}_{{\rm casc}}(Z_{s})}{\mathcal{G}_{{\rm casc}}(Z_{{\rm ns}})}=\frac{\left(1-\left|\Gamma_{s}\right|^{2}|\right)}{\left(1-\left|\Gamma_{{\rm ns}}\right|^{2}\right)}\frac{\left|1-S_{11}\Gamma_{{\rm ns}}\right|^{2}}{\left|1-S_{11}\Gamma_{s}\right|^{2}}\frac{\left|1-\Gamma_{{\rm rx}}\Gamma_{{\rm out}}(Z_{{\rm ns}})\right|^{2}}{\left|1-\Gamma_{{\rm rx}}\Gamma_{{\rm out}}(Z_{{\rm s}})\right|^{2}}.
\end{equation}
We now define a scale factor $\alpha$, which converts from power
to temperature units, and a mismatch factor $M_{s}$ (similar to Eq.
30 of \cite{Sutinjo:2020}): 
\begin{align}
\alpha & =\frac{T_{{\rm hot}}-T_{{\rm cold}}}{P_{{\rm hot}}-P_{{\rm cold}}}\\
M_{s} & =\left(1-\left|\Gamma_{{\rm ns}}\right|^{2}\right)\frac{\left|1-S_{11}\Gamma_{s}\right|^{2}}{\left|1-S_{11}\Gamma_{{\rm ns}}\right|^{2}}\frac{\left|1-\Gamma_{{\rm rx}}\Gamma_{{\rm out}}(Z_{s})\right|^{2}}{\left|1-\Gamma_{{\rm rx}}\Gamma_{{\rm out}}(Z_{{\rm ns}})\right|^{2}},\label{eq:mismatch}
\end{align}
and from Equation \ref{eq:power-basic} we retrieve
\begin{align}
\alpha P_{s}M_{s}-\left(1-\left|\Gamma_{s}\right|^{2}\right)\left(T_{s}+\frac{T_{{\rm rx}}}{\mathcal{G}_{{\rm DUT}}(Z_{s})}\right) & =\left(1-\left|\Gamma_{s}\right|^{2}\right)T_{n}.\label{eq:gain-cal-gamma}
\end{align}
Note that the terms $\mathcal{G}_{{\rm DUT}}^{-1}$ and $T_{{\rm rx}}$
are dependent upon the source impedance ($Z_{s}$); however, for sources
where $\left|\Gamma_{s}\right|\approx1$ and/or $\mathcal{G}_{{\rm DUT}}$
is large, the factor can be discarded, simplifying to: 
\begin{equation}
\alpha P_{s}M_{s}-\left(1-\left|\Gamma_{s}\right|^{2}\right)\left(T_{s}\right)\approx\left(1-\left|\Gamma_{s}\right|^{2}\right)T_{n}.
\end{equation}
We may now form the measurement vector $\v{t}'$ by applying
calibration Equation \ref{eq:gain-cal-gamma} to our measured power
$P_{s}$:
\begin{equation}
\v{t}'_{i}=\left(\alpha P_{s_{i}}M_{s_{i}}-\left(1-\left|\Gamma_{s_{i}}\right|^{2}\right)(T_{s_{i}}+\mathcal{G}_{{\rm DUT}}^{-1}(Z_{s})T_{{\rm rx}})\right).
\end{equation}
Specifically, if we connect a load, open, short, and a (lossless)
open 1/8-wavelength cable to a DUT, the noise parameters are retrieved
via
\begin{align}
\v{x} & =\text{(\ensuremath{\Amat^{\Gamma}}})^{-1}\left(\begin{array}{c}
\alpha P_{{\rm ld}}M_{{\rm ld}}-\left(1-\left|\Gamma_{{\rm ld}}\right|^{2}\right)(T_{{\rm amb}}+\frac{T_{{\rm rx}}}{\mathcal{G}_{{\rm DUT}}})\\
\alpha P_{{\rm op}}M_{{\rm op}}-\left(1-\left|\Gamma_{{\rm op}}\right|^{2}\right)(T_{{\rm amb}}+\frac{T_{{\rm rx}}}{\mathcal{G}_{{\rm DUT}}})\\
\alpha P_{{\rm sh}}M_{{\rm sh}}-\left(1-\left|\Gamma_{{\rm sh}}\right|^{2}\right)(T_{{\rm amb}}+\frac{T_{{\rm rx}}}{\mathcal{G}_{{\rm DUT}}})\\
\alpha P_{{\rm cbl}}M_{{\rm cbl}}-\left(1-\left|\Gamma_{{\rm cbl}}\right|^{2}\right)(T_{{\rm amb}}+\frac{T_{{\rm rx}}}{\mathcal{G}_{{\rm DUT}}})
\end{array}\right)\label{eq:noise-extract-full}\\
 & =\frac{1}{4}\left(\begin{array}{cccc}
4 & -1 & -1 & 0\\
0 & 0 & 1 & 0\\
0 & 1 & 0 & 0\\
0 & 1 & 1 & -2
\end{array}\right)\left(\begin{array}{c}
\alpha P_{{\rm ld}}M_{{\rm ld}}-(T_{{\rm amb}}+\frac{T_{{\rm rx}}}{\mathcal{G}_{{\rm DUT}}})\\
\alpha P_{{\rm op}}M_{{\rm op}}\\
\alpha P_{{\rm sh}}M_{{\rm sh}}\\
\alpha P_{{\rm cbl}}M_{{\rm cbl}}
\end{array}\right)\label{eq:noise-extract-simplified}
\end{align}
where $\text{(\ensuremath{\Amat^{\Gamma}}})^{-1}$ is given by Equation
\ref{eq:inverse}; note that $M_{{\rm ld}}\approx1$. 

\section{Measurement example\label{sec:Measurement-example}}

\begin{figure}
\begin{centering}
\includegraphics[width=0.95\columnwidth]{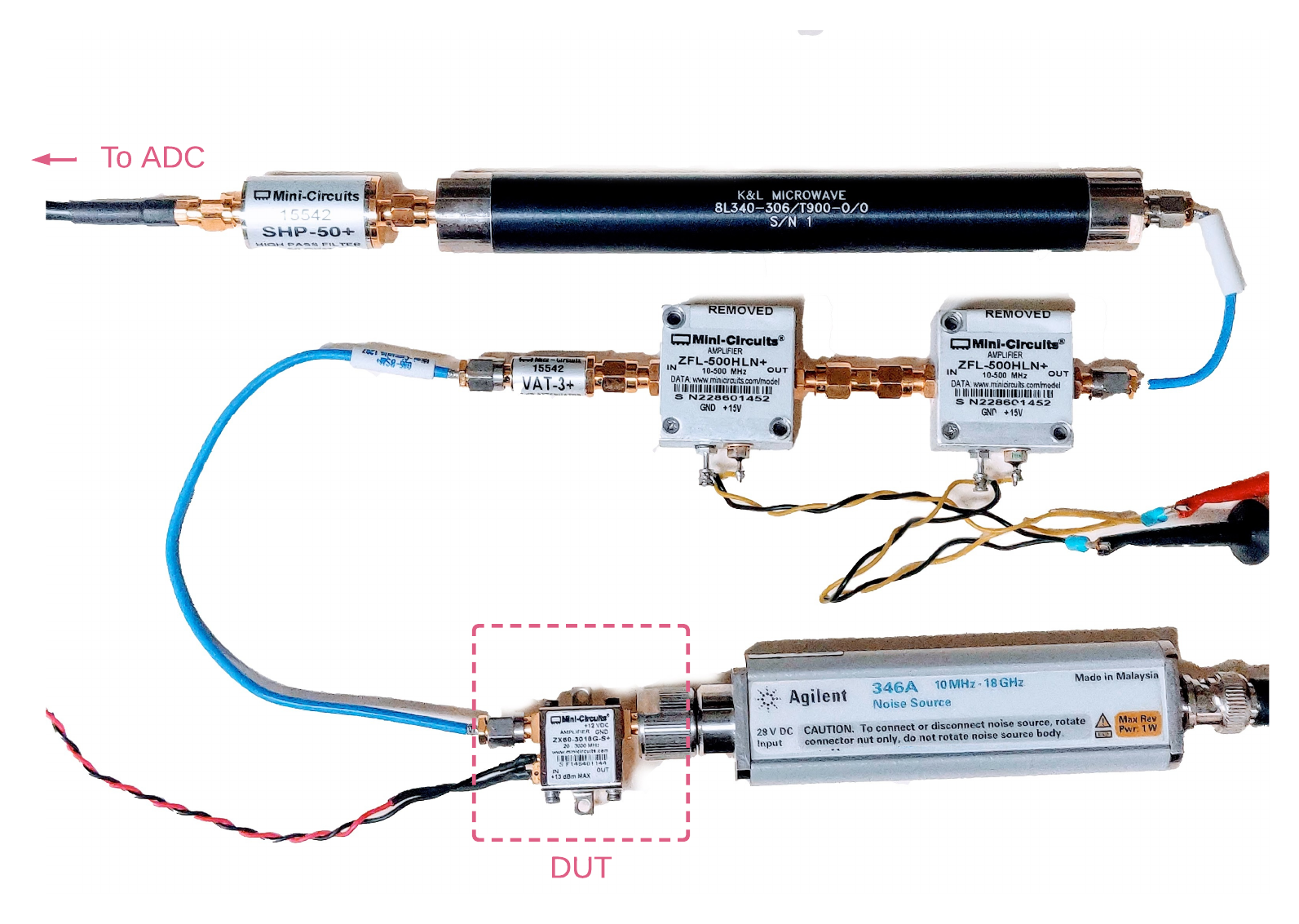}
\par\end{centering}
\caption{Example measurement setup as used to extract noise parameters. A DUT
(Minicircuits ZX60-3018G-S+) is connected to a simple noise receiver,
consisting of commercially-available amplifiers, filters, and cables
connected to an ADC. The receiver starts with a semi-rigid cable,
connected to a 3 dB attenuator to improve impedance matching. A pair
of Minicircuits ZFL-500HLN+ amplifiers provide $\sim38$ dB of gain,
and a K\&L Microwave lowpass filter and SHP-50+ highpass filter isolate
the 35\textendash 310 MHz band. In the photo, an Agilent 346A noise
source connected to the DUT. \label{fig:lab-setup} }
\end{figure}

\begin{figure}
\begin{centering}
\includegraphics[width=0.9\columnwidth]{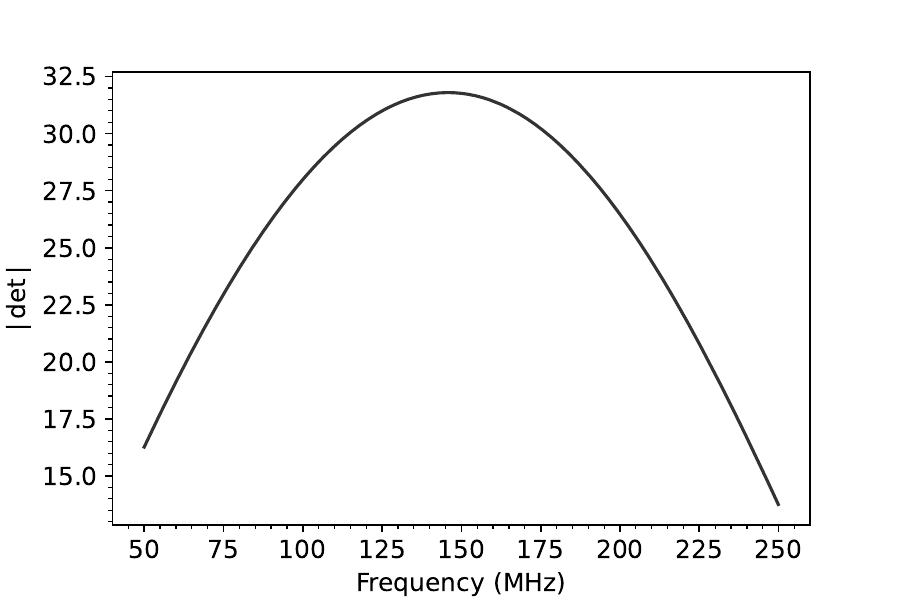}
\par\end{centering}
\caption{$|$det$|$ for the $4\times4$ $\Amat^{\Gamma}$ matrix formed from $\Gamma_{s}$
measurements of the reference sources, using an open 15-cm coaxial
cable. As the relative phase of $\Gamma_{{\rm cbl}}$ changes with
frequency, the $|$det$|$ is maximized at $\sim140$ MHz, close to the
ideal value of 32. \label{fig:abs-det-plot} }
\end{figure}

\begin{figure}
\begin{centering}
\includegraphics[width=0.9\columnwidth]{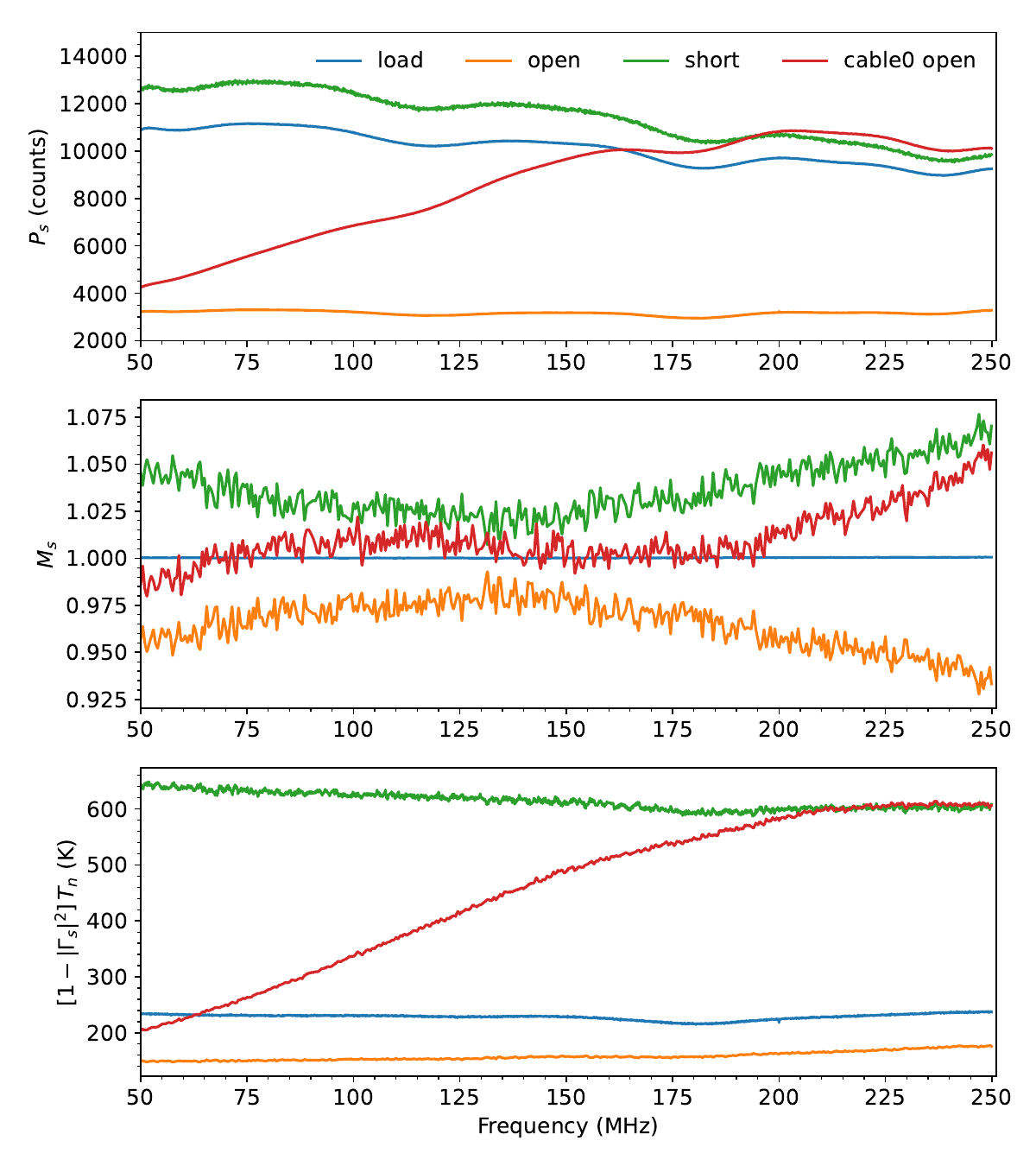}
\par\end{centering}
\caption{Measured power (uncalibrated) for the DUT (ZX60-3018G-S+) connected
to the load, open, short and cable references (top). Mismatch factors
$M_{s}$ applied during calibration (middle). Calibrated entries $(1-|\Gamma_{s_{i}}|^{2})T_{n}$
in the $\v{t}'$ vector (bottom).\label{fig:calibration-plot}}
\end{figure}

\begin{figure*}
\begin{centering}
\includegraphics[width=0.5\paperwidth]{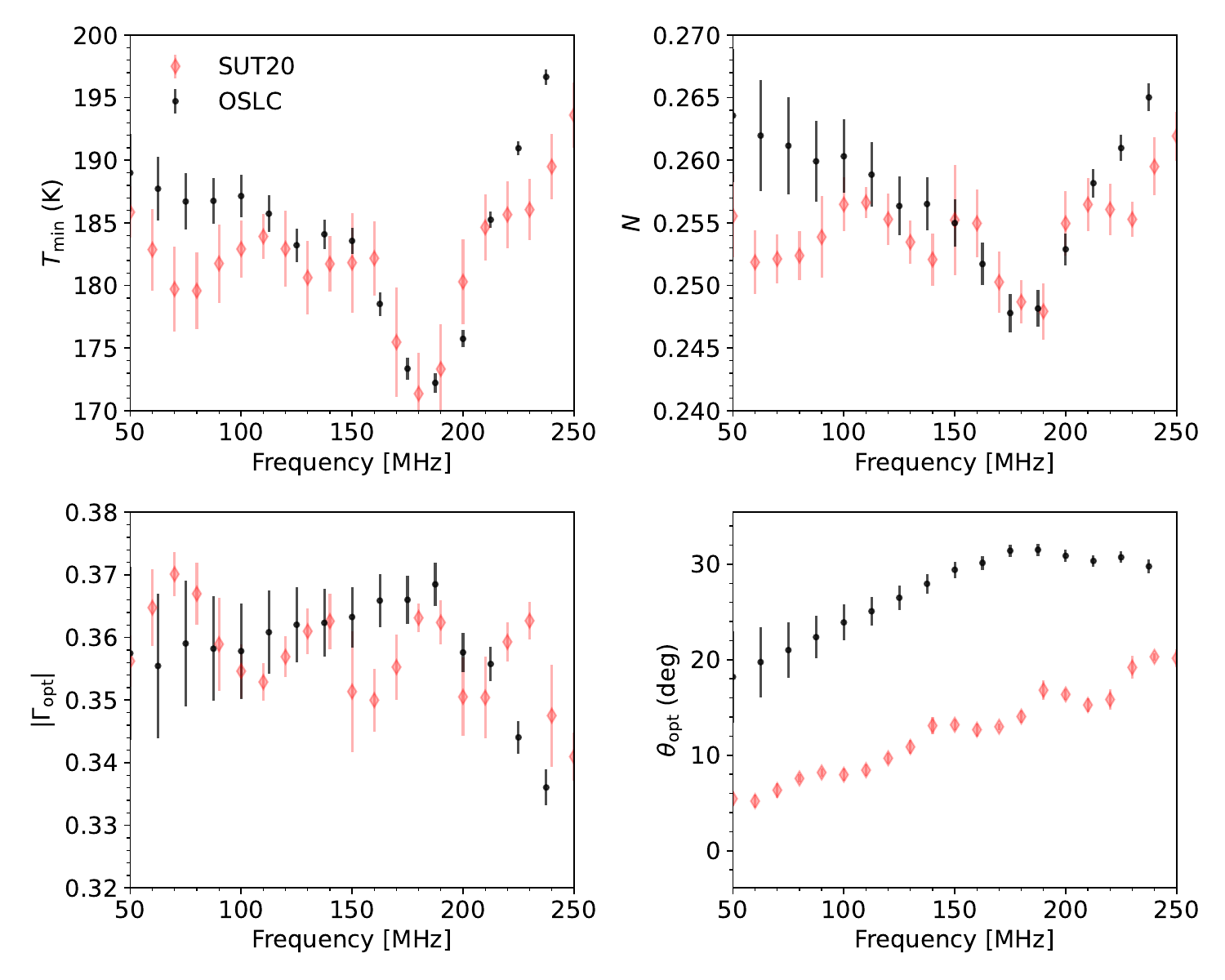}
\par\end{centering}
\caption{Noise parameters $T_{{\rm min}}$, $N$ , $|\Gamma_{{\rm opt}}|$ and  $\theta_{{\rm opt}}$
for the DUT (Minicircuits ZX60-3018G-S+) extracted using the method detailed in
Sec \ref{sec:method} (black); errorbars were determined using the Monte Carlo approach in \ref{subsec:Matrix-inversion-errors}. \ed{For comparison, the SUT20 noise parameter measurement for the same model amplifier (using a different technique and apparatus) are also plotted (red) \cite{Sutinjo:2020}.} \label{fig:Noise-parameters}}
\end{figure*}

We applied our technique to measure the noise parameters of a Minicircuits
ZX60-3018G-S+ amplifier, across 50\textendash 250\,MHz.  The amplifier
has a manufacturer-supplied noise figure of 2.7\textendash 2.9\,dB
across the 50\textendash 250\,MHz band, and a gain of $\sim25.5$
dB. Here, we used an SMA-terminated 15-cm RG-400 coaxial cable, with
$v_{c}\approx0.69.$ Based on Equations \ref{eq:freq_min} and \ref{eq:freq_max},
the nominal frequency range for this cable is $\sim$30\textendash 300\,MHz.
The $|$det$|$ for the $\Amat^{\Gamma}$ matrix using an open, short, load,
and 15-cm cable, is shown as a function of frequency in Fig \ref{fig:abs-det-plot}.

To generate hot and cold reference loads, we used a Keysight HP346A
noise source with an ENR of 5.56\textendash 5.49\,dB across 10\,MHz
to 1\,GHz. As ENR is only quoted at intervals, we use a 2-order polynomial
fit to generate values across 50\textendash 250\,MHz. 

Measurements of power spectra were generated using a custom receiver,
based on a 14-bit Signatek PX1500-2 analog to digital converter (ADC)
running at 650 Msamples/s; the noise performance of the receiver was
characterized prior to measurement. Power spectra were generated from
ADC samples via an autocorrelation spectrometer, which applies a 4096-channel
Fast Fourier Transform (FFT), signal detection, and time averaging. 
The receiver has a pair of Minicircuits ZFL-500HLN+ amplifiers to
provide an extra $\sim38$ dB of gain before digitization, and a 50-MHz
highpass filter (Minicircuits SHP-50+) and K\&L Microwave 300-MHz
lowpass filter were added at the ADC input as an anti-aliasing filter. 
\ed{The receiver temperature, $T_{\rm{rx}}$, is 1350--1450\,K across the band. 
We captured 60\,s of data per measurement. }

To measure reflection coefficients of the DUT, \ed{load, cable,}
and the receiver, we used a Fieldfox N9915A \ac{VNA},
calibrated with an Agilent 85052D calibration kit. Passive components
were measured using a high (0 dBm) port power, whereas lower power
(-30 dBm) was used to measure the {[}$S${]} matrix of the DUT and
$\Gamma_{{\rm rx}}$. Measurements were saved in S2P (Touchstone)
format, and read using the Python package \texttt{scikit-rf}\footnote{\href{https://www.scikit-rf.org}{www.scikit-rf.org}}. 
\ed{To determine $\Gamma_{{\rm sh}}$ and $\Gamma_{{\rm op}}$, we used a 
physical model based on the Agilent 85052D calibration standard definitions (see Appendix \ref{app:caldefs}).}

Figure \ref{fig:calibration-plot} shows the measure power $P_{s}$
for the DUT with the load, open short and cable reference sources
(top panel). The computed mismatch factors $M_{s}$ (Equation \ref{eq:mismatch})
which are applied during calibration (Equation \ref{eq:gain-cal-gamma})
are shown in the middle panel, and calibrated entries to the vector
$\v{t}'$ are shown in the bottom panel.

Based on the measurement vector $\v{t}'$ and $4\times4$ matrix
$\Amat^{\Gamma}$, we solve for the noise parameter vector $\v{x}$
(Equation \ref{eq:main-matrix-vector}) via matrix inversion $\v{x}=\text{(\ensuremath{\Amat^{\Gamma}}})^{-1}\v{t}'$
. We then solve for the standard noise parameters via Equations \ref{eq:np-fmin}\textendash \ref{eq:np-bopt}.
The derived noise parameters $T_{{\rm min}}$, $N$, and $\Gamma_{{\rm opt}}$
are shown in Figure \ref{fig:Noise-parameters}. Errors are derived
using the Monte Carlo approach detailed in Section \ref{subsec:Matrix-inversion-errors},
assuming VNA measurements of the coaxial cable are accurate to $\pm0.1$
dB in magnitude and $0.5^{\circ}$ in phase. Errors on the open, short
and load impedance are modelled as gaussian random variables, with
uncertainties based upon manufacturer-supplied electrical specifications
\cite{keysight:calkit}. 

\ed{Also plotted in Figure\,\ref{fig:Noise-parameters} are 
measurements from SUT20 \cite[Fig\,12]{Sutinjo:2020} of the same amplifier model. 
The authors followed a different measurement methodology, involving a Focus Microwaves CCMT-101 
single-probe slide screw tuner and Keysight PXA N9030A receiver. Note that 1) while the model is
identical, we did not use the same \ac{DUT} and 2) uncertainties
in SUT20 were derived using a different approach, so cannot be directly compared. Nevertheless,
our results are in close agreement with those presented in SUT20. A $\approx15^\circ$ offset between 
$\theta_{\rm{opt}}$ is apparent, which could be due to variation between amplifiers or differences
in operating conditions.}

\section{Discussion\label{sec:Discussion}}

Here, we have shown
that the commonly-used admittance-based matrix $\Amat^{G}$, after removal
of its singularity, can be rewritten as simple expression of reflection
coefficients ($\Amat^{\Gamma}$, Equation \ref{eq:AGamma}). 
We also show that uncertainties in noise parameter estimates
due to errors in the source impedance matrix \Amat\ are significantly
lowered by using $\Amat^{\Gamma}$ instead of $\Amat^{G}$. Combined, these
suggest that singularity-free formulations of \Amat\ should be used
where possible. 

We have presented a straightforward method to measure noise parameters
using a broadband load, open, short, and 1/8-wavelength cable as reference
sources (``OSLC''). Our method leverages a singularity-free matrix,
to allow the use of highly reflective reference loads. Specifically, our
method allows for open and short calibration standards to be used as 
reference sources. 

The ability to use highly reflective reference loads allows the spread
of loci in the Smith chart to be maximized, which also maximises $|$det$|$,
the magnitude of the matrix determinant. It follows that highly-reflective
references will yield lower worst-case scaling errors due to matrix
inversion. \ed{However, the use of highly-reflective sources requires 
that the \ac{DUT} is unconditionally stable.} \revtwo{Also, \ac{VNA} measurements of highly-reflective sources are prone to larger fractional errors (the magnitude of which depend upon \ac{VNA} specifications and calibration approach). Physical models of open and short circuits, as provided in \ac{VNA} calibration kits, can be used in lieu of \ac{VNA} measurements and are more accurate; methods for precise characterization of high-impedance references can also be employed \cite{dunsmore2011vna,Monsalve:2016}.}

The OSLC four-point method can be used across a range 0.2\textendash 1.8$\lambda_{0}$,
for a cable of length $\lambda_{0}/8$. If a larger range is required,
one could use a set of cables with varying lengths and repeat the
process at different central frequencies. Alternatively, rows may
be added to $\Amat^{\Gamma}$ and the pseudoinverse may be used when solving
(Equation \ref{eq:pseudoinverse}). 

The OSLC method is well-suited to low-frequency application ($<$1 GHz),
as the only source impedances are a 1/8-wavelength cable and a VNA
calibration kit: both are accessible and affordable. Approaches that
leverage rapid phase wrapping \cite{Hu:2004,rogers:2012,Roque:2021}
may require lengths of cable that are unwieldy or prohibitively long
at low frequencies. Admittance tuners, which are rated down to a half-wavelength,
are comparatively expensive and physically bulky. 

The OSLC method
is more practical than previous approaches for \emph{in-situ} measurement
in the field using a portable VNA and spectrum analyzer. On the flip
side, at higher frequencies, a 1/8-wavelength cable may become prohibitively
small. \ed{Precision airlines offer very low loss and 
may be a suitable alternative; nevertheless, further research is needed to
validate comparable methodologies at millimeter frequencies.} \revtwo{Modifications to the approach may also be needed for single-transistor DUTs, which can be poorly matched at low frequencies.}

\section*{Acknowledgments}

The authors would like to thank Budi Juswardy, Dave Kenney and Daniel Ung for
help with laboratory measurements. This research made use of Scikit-RF,
an open-source Python package for RF and Microwave applications, and
Corner.py \cite{cornerpy}. All calculations were performed in Python 3.7,
using the Numpy package \cite{harris2020array}.

\ifCLASSOPTIONcaptionsoff \newpage\fi

\bibliographystyle{IEEEtran}
\bibliography{IEEEabrv,references}

\appendices{}

\section{Derivation of $\Amat^{\Gamma}$\label{sec:Derivation-of}}

To show that ${\Amat}_{i}^{G'}=\Amat^{\Gamma}$, we note that 
\begin{align}
\left|\Gamma_{s}\right|^{2} & =\left(\frac{Y_{0}-Y_{s}}{Y_{0}+Y_{s}}\right)\left(\frac{Y_{0}-Y_{s}}{Y_{0}+Y_{s}}\right)^{*}\label{eq:gamma-squared}
\end{align}
where {*} denotes the complex conjugate, such that $Y_{s}^{*}=G_{s}-jB_{s}$.
Equation \ref{eq:gamma-squared} simplifies to:
\begin{equation}
\left|\Gamma_{s}\right|^{2}=\frac{Y_{0}^{2}-2G_{s}+(G_{s}^{2}+B_{s}^{2})}{Y_{0}^{2}+2G_{s}+(G_{s}^{2}+B_{s}^{2})}.\label{eq:gamma-squared2}
\end{equation}
Similarly, we may rewrite $\Gamma_{s_{I}}$ with the same denominator
by multiplying through by $1=(Y_{0}+Y_{s}^{*})/(Y_{0}+Y_{s}^{*})$:
\[
\Gamma_{s}=\left(\frac{Y_{0}-Y_{s}}{Y_{0}+Y_{s}}\right)\left(\frac{Y_{0}+Y_{s}^{*}}{Y_{0}+Y_{s}^{*}}\right)=\frac{Y_{0}^{2}-j2Y_{0}B_{s}+(G_{s}^{2}+B_{s}^{2})}{Y_{0}^{2}+2G_{s}+(G_{s}^{2}+B_{s}^{2})}.
\]
 from this, the following four quantities can be rewritten into terms
that share the denominator of Equation \ref{eq:gamma-squared2}:

\begin{align}
1-\left|\Gamma_{s}\right|^{2} & =\frac{4G_{s}}{Y_{0}^{2}+2G_{s}+(G_{s}^{2}+B_{s}^{2})}.\\
\left|1-\Gamma_{s}\right|^{2} & =\frac{4(G_{s}^{2}+B_{s}^{2})}{Y_{0}^{2}+2G_{s}+(G_{s}^{2}+B_{s}^{2})}\\
\left|1+\Gamma_{s}\right|^{2} & =\frac{4Y_{0}^{2}}{Y_{0}^{2}+2G_{s}+(G_{s}^{2}+B_{s}^{2})}\\
{\rm Im}(\Gamma_{s}) & =\frac{-2Y_{0}B_{s}}{Y_{0}^{2}+2G_{s}+(G_{s}^{2}+B_{s}^{2})}.
\end{align}
Using these three identities 
\begin{align}
\frac{\left(1-\left|\Gamma_{s}\right|^{2}\right)}{G_{s}} & =\frac{\left|1+\Gamma_{s}\right|^{2}}{G_{0}}\\
\left(1-\left|\Gamma_{s}\right|^{2}\right)\frac{\left|Y_{s}\right|^{2}}{G_{s}} & =G_{0}\left|1+\Gamma_{s}\right|^{2}\\
\left(1-\left|\Gamma_{s}\right|^{2}\right)\frac{B_{s_{i}}}{G_{s_{i}}} & =-2{\rm Im}(\Gamma_{s})
\end{align}
We thus have that $\Amat^{G'}=\Amat^{\Gamma}$ (setting $Y_{0}=G_{0}=1$). 

\section{Reverse relations\label{sec:Reverse-relations}}

The reverse relations for Equations \ref{eq:np-fmin}\textendash \ref{eq:np-bopt}
are:

\begin{align}
a & =T_{{\rm min}}-2R_{N}T_{0}G_{{\rm opt}}\label{eq:a-rev}\\
b & =R_{N}T_{0}Y_{0}\\
c & =\frac{R_{N}T_{0}}{Y_{0}}(G_{{\rm opt}}^{2}+B_{{\rm opt}}^{2})\\
d & =-2R_{N}T_{0}B_{{\rm opt}}.\label{eq:d-rev}
\end{align}
These reverse relations are used in Section \ref{sec:A-reflection-coefficient}
to simulate noise parameter measurements.

\ed{\section{Open and short circuit models}\label{app:caldefs}}
\ed{Open and short source impedances are physically modelled by a line terminated with an inductance (short) or capacitance (load) (see \cite{dunsmore2011vna,DeGroot:2000}). 
Following \cite{agilentCalStandard}, the reflection coefficient of a terminated line is given by 
\begin{equation}
\Gamma_{i}=\frac{\Gamma_{1}\left(1-e^{-2\gamma\ell}-\Gamma_{1}\Gamma_{T}\right)+e^{-2\gamma\ell}}{1-\Gamma_{1}\left[e^{-2\gamma\ell}\Gamma_{1}+\Gamma_{T}\left(1-e^{-2\gamma\ell}\right)\right]}
\end{equation}
where $\Gamma_1$ is the transmission line reflection coefficient, $\Gamma_T$ is the impedance of the termination, $\ell$ is the transmission line length, $\gamma$ is the propagation constant along the transmission line. $\Gamma_1$ is related to $Z_c$, the characteristic impedance of the line, by
\begin{equation}
\Gamma_1 = \frac{Z_c - Z_r}{Z_c + Z_r}.
\end{equation}
Calibration kits provide a table of calibration standard definitions, from which $Z_c$ can be determined for a given frequency. 
Specifically, the manufacturer provides an offset delay $\tau_{\rm{ofs}}$ (in ps), offset loss at 1\,GHz $\l_{{\rm ofs}}$ (G$\Omega$/s), and offset impedance $Z_{{\rm ofs}}$, from which $Z_c$ and $\gamma\ell$ are found:
\begin{align}
Z_{C}	&=\left(Z_{{\rm ofs}}+\frac{Z_{{\rm ofs}}}{2\omega}\sqrt{f_{{\rm GHz}}}\right)-j\left(\frac{l_{{\rm ofs}}}{2\omega}\sqrt{f_{{\rm GHz}}}\right) \\
\gamma\ell	&=\left(\frac{l_{{\rm ofs}}\tau_{{\rm ofs}}}{2Z_{{\rm ofs}}}\sqrt{f_{{\rm GHz}}}\right)+j\left(\omega\tau_{{\rm ofs}}+\frac{l_{{\rm ofs}}\tau_{{\rm ofs}}}{2Z_{{\rm ofs}}}\sqrt{f_{{\rm GHz}}}\right)
\end{align}
In the Agilent 85052D calibration kit  \cite{keysight:calkit}, 
the inductance model for a short circuit is a third-order polynomial in frequency
\begin{align}
L_{\rm{sh}}(f) &= L_0 + L_1 f + L_2 f^2 +L_3 f^3 \\
Z_{\rm{sh}}    &= j 2 \pi f L_{\rm{sh}}(f).
\end{align}
Similarly, the capacitance model for an open circuit is 
\begin{align}
C_{\rm{op}}(f) &= C_0 + C_1 f + C_2 f^2 + C_3 f^3 \\
Z_{\rm{op}}    &= -j (2 \pi f)^{-1}.
\end{align}
From these equations, a model for $\Gamma_{\rm{op}}$ and $\Gamma_{\rm{sh}}$ may be formed. Calibration standard definitions for the open and short are summarized in Tab.\,\ref{tab:caldef}.

The 85052D open and short have electrical specifications of $\pm0.65^\circ$ and  $\pm0.50^\circ$ deviation from nominal, respectively, at DC to 3\,GHz. 

\begin{table}[h]
\centering
\caption{Agilent 85052D open/short calibration definitions}\label{tab:caldef}
\begin{tabular}{clc}
\hline
\hline
\multicolumn{3}{c}{Open} \\
\hline
Offset delay & 29.243 & ps \\
Offset loss  & 2.2 & G$\Omega$/s \\
$C_0$  & $49.433\times10^{-15}$ & F         \\
$C_1$  & $-310.13\times10^{-27}$ & F/Hz       \\
$C_2$  & $23.168\times10^{-36}$ & F/Hz$^2$   \\
$C_3$  & $-0.15966\times10^{-45}$ & F/Hz$^3$   \\
\hline
\multicolumn{3}{c}{Short} \\
\hline
Offset delay & 31.785 & ps \\
Offset loss  & 2.36  & G$\Omega$/s \\
$L_0$ & $2.0765\times10^{-12}$ & H  \\
$L_1$ & $-108.54\times10^{-24}$ & H/Hz  \\
$L_2$ & $2.1705\times10^{-33}$ & H/Hz$^2$  \\
$L_3$ &$-0.01\times10^{-42}$ & H/Hz$^3$  \\
     
\hline
\multicolumn{3}{r}{Offset impedance $Z_{\rm{ofs}}$=50\,$\Omega$ }\\
\end{tabular}

\end{table}
} 

\end{document}